\documentclass{aa}
\usepackage{graphicx}
\usepackage{rotate}
\usepackage{epsf}
\newcommand{\DD}{\frac}

\newcommand{\ber}{\begin{array}}
\newcommand{\eer}{\end{array}}

\newcommand{\lra}{\longrightarrow}

\newcommand{\Msun}{{\,{\cal M}_{\odot}}}
\newcommand{\Mdot}{{\,\dot{\cal M}}}
\begin{document}
  \title{Ion-dominated plasma and the origin of jets in quasars }

   \author{
              A. Hujeirat${}^{1}$,
              M. Camenzind${}^{2}$,
              M. Livio${}^{3}$
          }


   \offprints{A. Hujeirat}

   \institute{${}^{1}$Max-Planck-Institut f\"ur Astronomie, D-69117 Heidelberg, Germany \\
              ${}^{2}$Landessternwarte-Koenigstuhl, D-69117, Heidelberg, Germany \\
     ${}^{3}$Space Telescope Science Institute, 3700 San Martin Drive, Baltimore, MD 21218, USA
             }
   \date{Received  ***;  \underline{ACCEPTED 03/09/2002}}

   \abstract{  Low cooling plasmas associated with large kinetic
             energies are likely to be the origin of the kpc-extended and well collimated
             extra-galactic jets.

             It is proposed that jets are launched  from a layer, governed by a
             highly diffusive, super-Keplerian rotating  and thermally dominated by
             virial-hot and magnetized ion-plasma. The launching layer is located between the
             accretion disk and the corona surrounding the nucleus. 
             The matter in the layer
             is causally connected to both the disk and to the central engine.
             Moreover we find that coronae, in the absence of heating from
             below,  are  dynamically unstable to thermal ion-conduction,
             and that accretion disks  become intrinsically advection-dominated.

             We confirm the capability of this multi-layer model to form jets by 
             carrying out 3D axisymmetric
             quasi-stationary MHD calculations with high spatial resolution, and taking into account
             turbulent and magnetic diffusion. 
             The new multi-layer topology accommodates several previously proposed elements for 
             jet-initiation, in particular the ion-torus,  the magneto-centrifugal and 
             the truncated disk - advective tori models.

   }

   \titlerunning{Ion-dominated plasma -  origin of jets}
   \maketitle
%
\section{Introduction}
\label{intro}
Jets have been observed in many systems including active galaxies, X-ray binaries,
black holes X-ray transients, supersoft X-ray sources and young stellar objects
(K\"onigl 1997, Livio 1999, Mirabel 2001).
Each of these systems is considered to contain an accretion disk, while 
jet-speeds have been verified to be of the order of the escape velocity at the
vicinity of the central object (Mirabel 1999, Livio 1999 and the references therein).
Recent observations of the M87 galaxy reveal a significant jet-collimation 
already at 100 gravitational radii from the central engine, and that jet-launching
should occur close to the last stable orbit (Biretta et al. 2002). 

Several scenarios have been suggested to uncover the mechanisms underlying 
jet-initiations and their connection to accretion disks (Pudritz \& Norman 1986). 
In most of these models magnetic
fields (-MFs) are considered to play
the major role in powering and collimating jets (e.g., the magneto-centrifugal acceleration
model of Blandford \& Payne 1982, the ion-torus model of Rees et al. 1982, X-point
model of Shu et al. 1995, ADAF and ADIOS models of Narayan \& Yi 1995 and
Blandford \& Begelman 1999).

Previous radiative hydrodynamical studies without magnetic fields have confirmed
 the formation of a transition 
layer (-TL) between the disk and the corona, governed by thermally-induced 
outflows (Hujeirat \& Camenzind 2000).
The aim of this paper is show that incorporating large scale magnetic fields (-MFs)
manifests such formation and dramatically strengthen the dynamic of the in- and
out-flows. Moreover, the TL is shown to be an optimal runaway region 
where highly energetic ion-jets start off. The back reaction of jet-flows 
on the structure of the disk and on the corona surrounding the nucleus is investigated also.

The study is based on self-consistent 3D axi-symmetric quasi-stationary MHD calculations, 
taking into account magnetic and hydro-turbulent diffusion, and adopting the two-temperature
description (Shapiro et al. 1976). 
This adaptation is fundamental as 1) the dynamical time scale
around the last stable orbit may become shorter than the Coulomb-coupling time. Therefore
turbulent dissipation, adiabatic or shock compression preferentially heat up the ions rather than
electrons ($T_\mathrm{i} \propto \rho^{2/3}_\mathrm{i}$, while 
$T_\mathrm{e} \propto \rho^{1/3}_\mathrm{i}$).
2) Taking into account that ions radiate inefficiently, having virial-heated ions 
    in the vicinity of the last stable orbit
   is essential for the total energy-budget of large scale jets.
\begin{figure}[htb]
\centering
{\hspace*{-0.1cm}
\includegraphics*[width=8.0cm, bb=0 0 392 194,clip]
{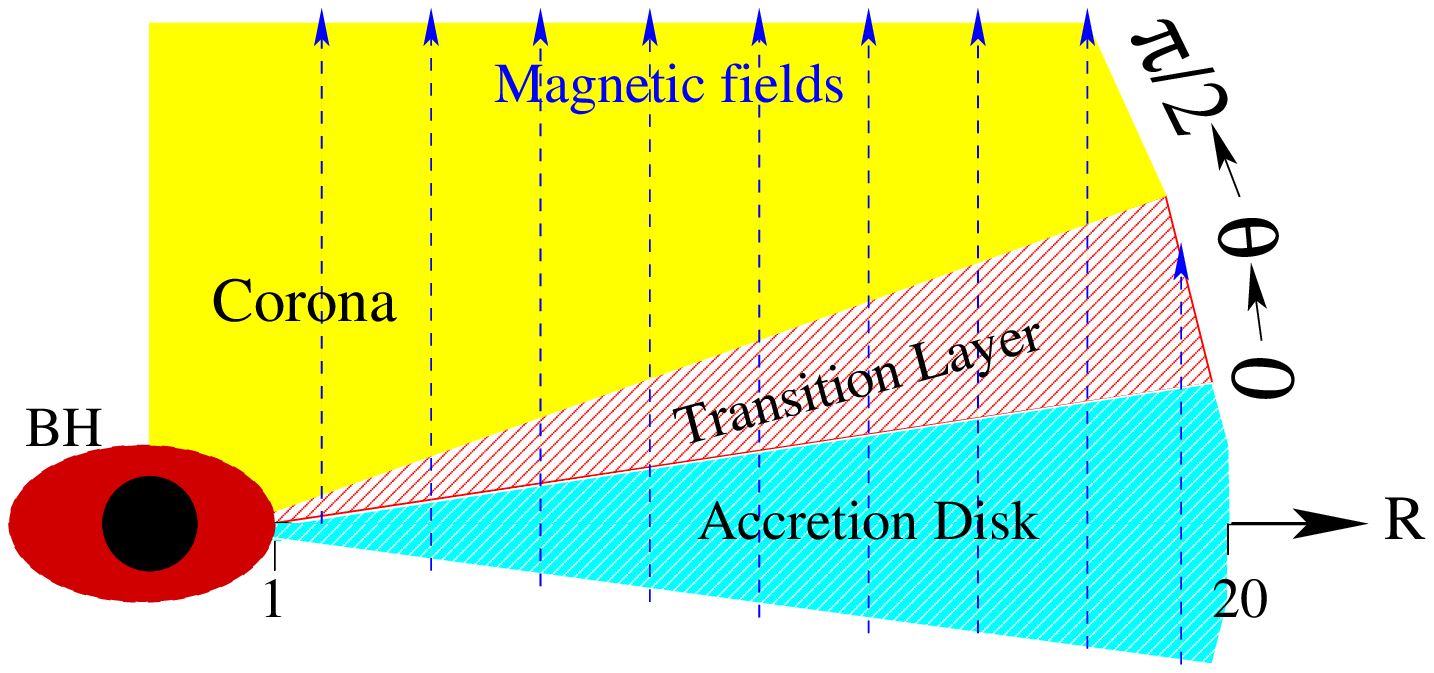}
}
{\vspace*{-0.1cm}}
\caption{
The model consists of  $10^8M_{\odot}$ Schwarzschild BH at the center
 (its gravity
is described in terms of the quasi--Newtonian potential of Paczynski\,\&\,Wiita 1980),
and an  SS-disk (blue color, extending from r=1 to r=20 in units of the radius of the
last stable orbit i.e., in $3\times R_\mathrm{Schwarzschild}$, thickness
 $H_\mathrm{d} = 0.1 r,$ an accretion rate of
 $\Mdot = 0.01\times \Mdot_\mathrm{Edd},$ and a central disk
temperature of $T=10^{-3} T_\mathrm{virial}$ at the outer radius). The ion-temperature
  $\rm T_\mathrm{i}$  is set to be equal to the electron temperature 
$\rm T_\mathrm{e}$ initially.) 
The low-density hot corona
($T=T_\mathrm{virial}$, and density $\rho(t=0,r,\theta)=10^{-4}\rho(t=0,r,\theta=0$)
is set to envelope the disk.
A large scale magnetic field is set to thread the disk and the overlying corona
(blue lines, $\beta = P_\mathrm{mag}/P_\mathrm{gas} = 1/4$ at the outer radius,
where $P_\mathrm{gas}$
is the ion-pressure,  $P_\mathrm{mag} = B\cdot B/8 \pi$ is the magnetic pressure, and
$\rm B$ is the magnetic field whose components are
 $(B_\mathrm{P}, B_\mathrm{T})= (B1,B2,B_\mathrm{T})$.)
 The low-density hot corona 
($T=T_\mathrm{virial}$, and density $\rho(t=0,r,\theta)=10^{-4}\rho(t=0,r,\theta=0$)
is set to envelope the disk.
A large scale magnetic field is set to thread the disk and the overlying corona
(blue lines, $\beta = P_\mathrm{mag}/P_\mathrm{gas} = 1/4$ at the outer radius, where $P_\mathrm{gas}$
is the ion-pressure,  $P_\mathrm{mag} = B\cdot B/8 \pi$ is the magnetic pressure, and
$\rm B$ is the magnetic field whose components are
 $(B_\mathrm{P}, B_\mathrm{T})= (B1,B2,B_\mathrm{T})$.) 
The numerical procedure is based on using the implicit solver IRMHD3  to search steady-state
solution for the 3D axi-symmetric two-temperature diffusive MHD equations in spherical
geometry (for further clarifications about the equations and the numerical method
 see Hujeirat \& Rannacher 2001, Hujeirat \& Camenzind 2000). 
The ion-pressure is used to describe the turbulent viscosity:
$\nu_\mathrm{turb} = \alpha P_\mathrm{gas}/\Omega$, where $\alpha$ is the usual viscosity
 coefficient, and $\Omega$ is
the angular frequency. The magnetic diffusivity
is taken to be equal to $\nu_\mathrm{turb}.$
$250\times80$ strongly stretched finite volume cells in the radial and vertical direction
have been used. Normal symmetry and anti-symmetry boundary conditions have been imposed along
the equator and the rotation axis. Extrapolation has been adopted to fix down-stream values
at the inner boundary. Non-dimensional formulation is adopted, using the reference scaling variables:
$\rm{\tilde{\rho}= 2.5 \times 10^{-12} {\rm g\, cm^{-3}}, \tilde{T}= 5 \times 10^7 K},$
$\tilde{U}=\tilde{V_\mathrm{S}} = \gamma {\cal R}_\mathrm{g} \tilde{T}/
\mu_\mathrm{i},$ $(\mu_\mathrm{i}=1.23)$.
$\tilde{B}=\tilde{V_\mathrm{S}} \sqrt{4 \pi \tilde{\rho}}.$ 
The location of the transition layer (-TL), where the ion-dominated plasma
is expected to rotate super-Keplerian and being accelerated into jets,
is shown for clarity. 
 }
\end{figure}
\section{Formation of the super-Keplerian layer}
Angular momentum in standard disks (Shakura \& Sunyaev 1973, hereforth SS-disks)
is transported outwards mainly via small scale magneto-hydrodynamical
turbulence. Magnetic fields (-MFs) however were assumed to be weak, and remain so during
the whole viscous evolution of the disk.
Here we adopt a different view. Let
 $\nu_\mathrm{tot}= \nu_\mathrm{HD} + \nu_\mathrm{Mag} = \alpha (P_\mathrm{gas} + \pi_\mathrm{B})/\Omega$
be a modified  dynamical viscosity (see the caption of Figure 1 for elaboration). 
 Unlike $\nu_\mathrm{HD}$ which transports angular momentum outwards, 
$\nu_\mathrm{Mag} $ transports angular momentum in the vertical direction, provided that 
$ B_\mathrm{P}$ is of large scale topology. Taking into account that  $\nu_\mathrm{HD}$ in SS-disks
decreases inwards,  a transition radius $R_\mathrm{Tr}$ at which 
{result}\footnote{If $\nu_\mathrm{Mag}$ is too small, then magnetic fields 
are frozen-in to the gas approximately. In this
  case  $B \sim \rho^{2/3}$, and B must increase inwards, irrespectively, whether the disk is standard
  or advection-dominated}.
the time scale of angular momentum removal
by MFs at $R_\mathrm{Tr}$  is of the same order as that by the turbulent viscosity. Hence \\
\begin{equation}
\DD{\tau_\mathrm{rem}}{\tau_{tur}} \simeq \alpha (\DD{H}{r})(\DD{V_\mathrm{s}}{V_\mathrm{A}})^2 \simeq 1,
\end{equation}
where ${V_\mathrm{s}},~V_\mathrm{A}$ are the sound and transverse $\rm Alf\grave{v}en$ speeds
$(V_\mathrm{A}=\sqrt{B_\mathrm{P} B_\mathrm{T}/4 \pi \rho})$, respectively.
Taking into account that $ {H}/{r} \ge \alpha$, we end up with  ${V_\mathrm{A}} \ge \alpha{V_\mathrm{s}}.$
The last optimistic inequality applies everywhere in the disk, irrespective whether
 turbulence is mediated by $\pi_\mathrm{B}$
or not. Therefore, vertical transport of angular momentum via MFs
is at least as efficient as $\alpha-$viscosity. This is even more justified by the fact that
Balbus-Hawley instability in Keplerian-disks
amplifies initially weak fields to considerably large values,
but remain still  below  equipartition (Hawley et al. 1996). Imposing appropriate 
boundary conditions, and carrying the MHD-Box calculations with high spatial 
resolution, MFs could be amplified up to equipartition (Ziegler 2002), yielding thereby
${\tau_\mathrm{rem}} < {\tau_{tur}}$.
Furthermore, the dynamo-action model proposed by Tout \& Pringle 1992, if applied,  would make 
${\tau_\mathrm{rem}}$ even shorter.

To be noted here that when taking a more realistic density and temperature stratification 
in global 3D MHD disk-calculations, vertical transport of angular momentum is inevitable (Arlt 2002).

Efficient vertical transport of angular momentum rises the following important issues:
1) Turbulence in the disk need not be dissipated, or it might be even  suppressed by the
   amplified $B_\mathrm{P}$. This allows accretion to evolve without necessarily 
   emitting the bulk of the their potential energy as radiation, and gives rise to
   energy re-distribution.
   2) Accretion flows may not proceed as slowly as in
   SS-disk, but they may turn into 
advection-{dominated}\footnote{The flow is said to be advection-dominated if transport via fluid-motion
 dominates viscous re-distribution.}.
This occurs  because
the time scale of angular momentum removal 
from the disk scales as: 
\begin{equation}
  \tau_\mathrm{rem} \sim \rho V_\mathrm{T} H/B_\mathrm{P} B_\mathrm{T} \sim r^{3/2},
\end{equation}
where $V_\mathrm{T} (=r \cos{\theta}~ \Omega)$ is the angular velocity (Fig. 1). 
This indicates that angular momentum removal is more efficient at smaller radii.

To maintain dynamical stability, angular 
momentum removal from the disk should be compensated by rapid advection from larger radii,
 i.e., $\tau_\mathrm{adv} =\tau_\mathrm{rem}$.
Further, rapid and steady generation of $B_\mathrm{T}$ in the disk yields
$B_\mathrm{P}/B_\mathrm{T} = H_\mathrm{d}/r.$ The later two conditions imply that the radial velocity $U_\mathrm{r} \sim V_\mathrm{A}$, 
which means that the stronger the MF threading the disk, the more advection-dominated
it becomes,  and therefore the faster is the establishment of the super-Keplerian layer. 
 
Based on the present calculations (see the caption of Figure 1 and figures 2- 4), it is found that:   1) Angular velocity  
in the transition layer (TL) adopts approximately the profile
$\Omega \sim r^{-5/4}$. 2) Energy dissipation is injected primarily 
into the ions that cool predominantly through fast outflows. 
3) The generated 
toroidal magnetic field is quenched by a magnetic diffusion (reconnection)
and fast outflows. 
The width of the TL is pre-dominantly determined through the transverse
variation of the ion-pressure across the jet, i.e.,  $ H_\mathrm{W} = 
P_\mathrm{i}/\nabla P_\mathrm{i} \approx 0.2\,r$,
and so strongly dependent on whether the flow is a one- or two-temperature plasma.
In the steady-state case, this implies:
\begin{equation}
   \rho   \sim r^{-7/4},
   T_\mathrm{i}  \sim r^{-1/2},
   U_\mathrm{r}  \sim r^{-1/4},
   B_\mathrm{P}/B_\mathrm{T} \sim \rm const.
\end{equation}

We note that  $U_\mathrm{r}$ adopts a profile and attains values similar  to those 
in the innermost part of the disk.
Provided that energy exchange between the matter
in the disk and in the TL is efficient, the incoming matter can easily  be
re-directed into outwards-oriented motions. This implies that the Bernoulli number (Be) 
can change sign in dissipative flows. As Fig. 4 shows, 
Be is  everywhere negative save the TL, where it attains large positive
values, so that the ion-plasma can start its kpc-journey.
\begin{figure}[htb]
\centering
{
\includegraphics*[width=7.5cm, bb=42 274 315 740,clip]
{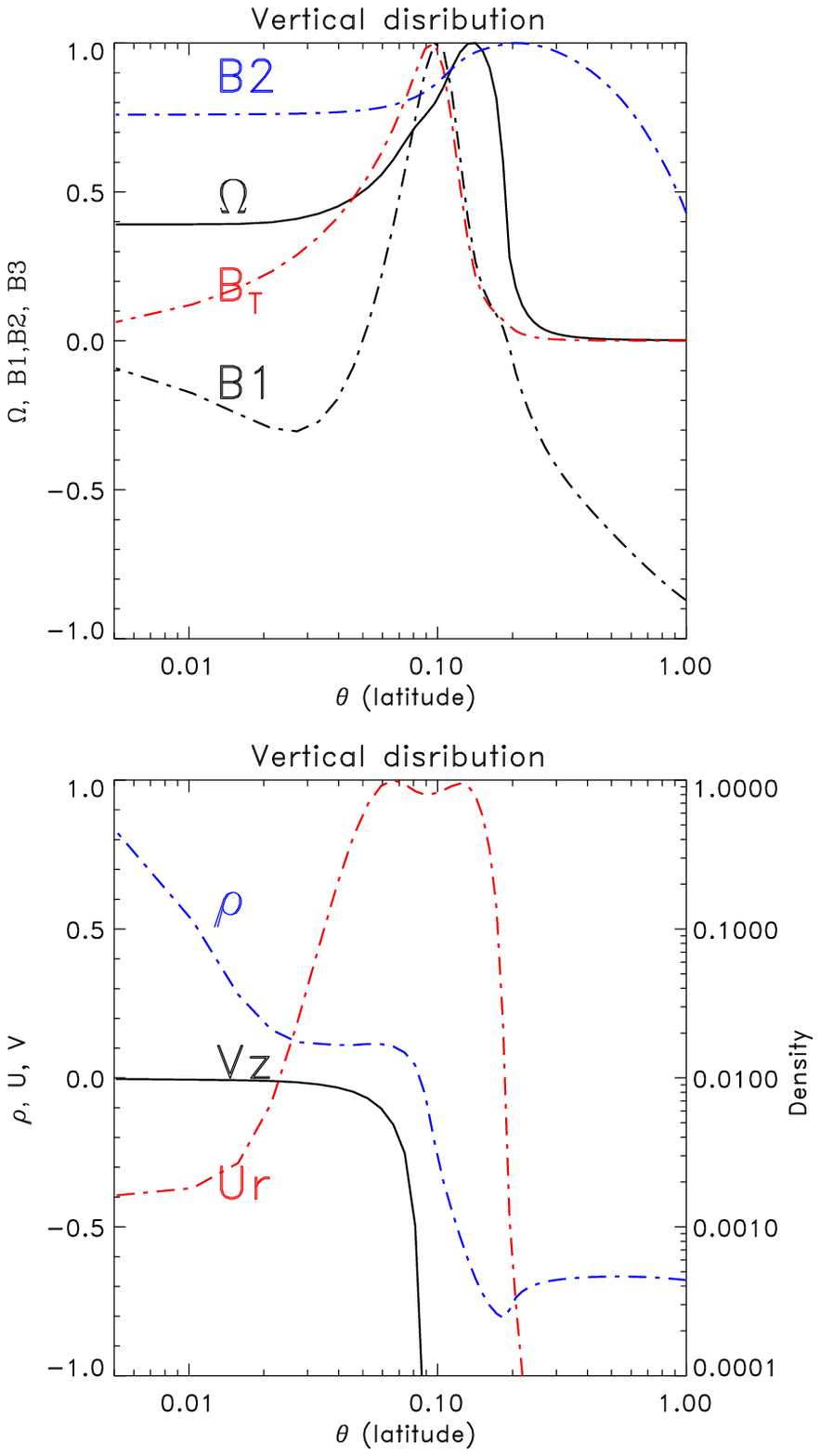}
}
\caption{ The horizontal distribution of the normalized density $\rho$, 
angular velocity $\Omega$, radial and horizontal velocities 
$U_\mathrm{r},\,V_\mathrm{z},$ the radial and horizontal
MF-components $(B1, B2)=(B_\mathrm{r}, B_\theta)=B_\mathrm{P}$ and the toroidal
MF-component $B_\mathrm{T}$ at $r=2.5$. Note the density plateau, 
the positive radial velocity (outflows), the  super-Keplerian rotation and
the strongly enhanced strength of the MF-components in the TL. The inwards-oriented
motions in the disk (inflows) and strongly increasing density towards the equator 
are obvious. }
\end{figure}
Worth-noting is the resulting MF topology. Apparently, the outflow
is sufficiently strong to shift the MF lines outwards, while the large
diffusivity prevents the formation of large electric currents along the equator. In the corona
however, MFs are too weak to halt the diffusive plasma in the dynamically unstable
corona against gravity, and instead, they drift with the
infalling gas inwards. In the case of very weak MFs ($\beta \le 0.1$),  
our calculations indicate a considerably weak outflow.  This is a consequence of the tendency of
the MFs to establish a monopole like-topology, i.e, a one-dimensional MF-topology in which 
$\rm{B}_\theta \lra 0$.
In this case, the magnetic tension $\pi_\mathrm{B}$  becomes inefficient in feeding  the matter
in the TL with the angular momentum required for launching jets, indicating herewith that
cold accretion disks alone are in-appropriate for initiating winds (Ogilvie \& Livio 2001).

Comparing the flux of matter in the wind to that in the disk, we find that
${\Mdot}_\mathrm{W}/{\Mdot}_\mathrm{d} = {\rm const.} \approx 1/20$. The angular momentum flux associated with
the wind  is
 $\dot{\cal J}_\mathrm{W}/\dot{\cal J}_\mathrm{d} = {\rm a }\, 
({\Mdot}_\mathrm{W}/{\Mdot}_\mathrm{d})\,r^{1/4},$ where ${\rm a }$ is a constant of order unity.
Consequently, at $r = 3\times 10^2 R_\mathrm{Sch}$,  almost $25\%$ of the total accreted angular momentum
in the disk re-appears in the wind. \\ \\
Why is the TL  geometrically thin?\\
In stratified dissipative flows the density scale height is much smaller than 
the scale height of the angular velocity (Hujeirat \& Camenzind 2000). Since the flow
in TL rotates super-Keplerian, Coriolis forces act to compress the
disk-matter and make its density scale height even smaller. This implies that advective-disks
are geometrically thin, much thinner  than what ADAF-solutions predict. 
\\
On the other hand, unless there is a significant energy flux that heats up the plasma 
from below, as in the case of stars, heat conduction will always force the BH-coronae
to collapse dynamically.  
To elaborate this point, let us compare the conduction time scale with the dynamical
time scale along $\rm B_\mathrm{P}$-field at the last stable orbit of a 
SMBH:
\begin{equation}
\DD{\tau_{\rm cond}}{\tau_{\rm dyn}} =  \DD{r \rho U_\mathrm{r}}{\kappa_0 T^{5/2}_\mathrm{i}}
 = 4.78\times 10^{-4} \rho_{10} T^{-5/2}_\mathrm{i,10} {\cal M}_8, 
\end{equation}
where $\rho_{10}$, $\rm{T_\mathrm{i,10}}$ and ${\cal M}_8$ 
 are respectively in $10^{-10}\,\rm{g\,cm^{-3}}$, 
 $10^{10}\,$K and in $10^8\,\Msun$ units.
This is much less than unity for most reasonable values of density and temperature typical for 
AGN-environments. In writing Eq. 4 we have taken optimistically the upper limit $ c/\sqrt{3}$ for the velocity,
and set $\kappa_0 = 3.2\times 10^{-8}$ for the ion-conduction coefficient.
When modifying the conduction operator to respect causality, we obtain 
${\tau_{\rm cond}}/{\tau_{\rm dyn}} \le U_\mathrm{r}/c $, which is again smaller than unity.
 \\
This agrees with our numerical calculations which rule out the 
possibility of outflows  along the rotation axis, and in particular not 
from the highly unstable polar region of the BH, as ADAF-solutions predict.
\begin{figure*}
\begin{center}
{\hspace*{-0.2cm}
\includegraphics*[width=13.5cm, bb=46 270 515 740,clip]
{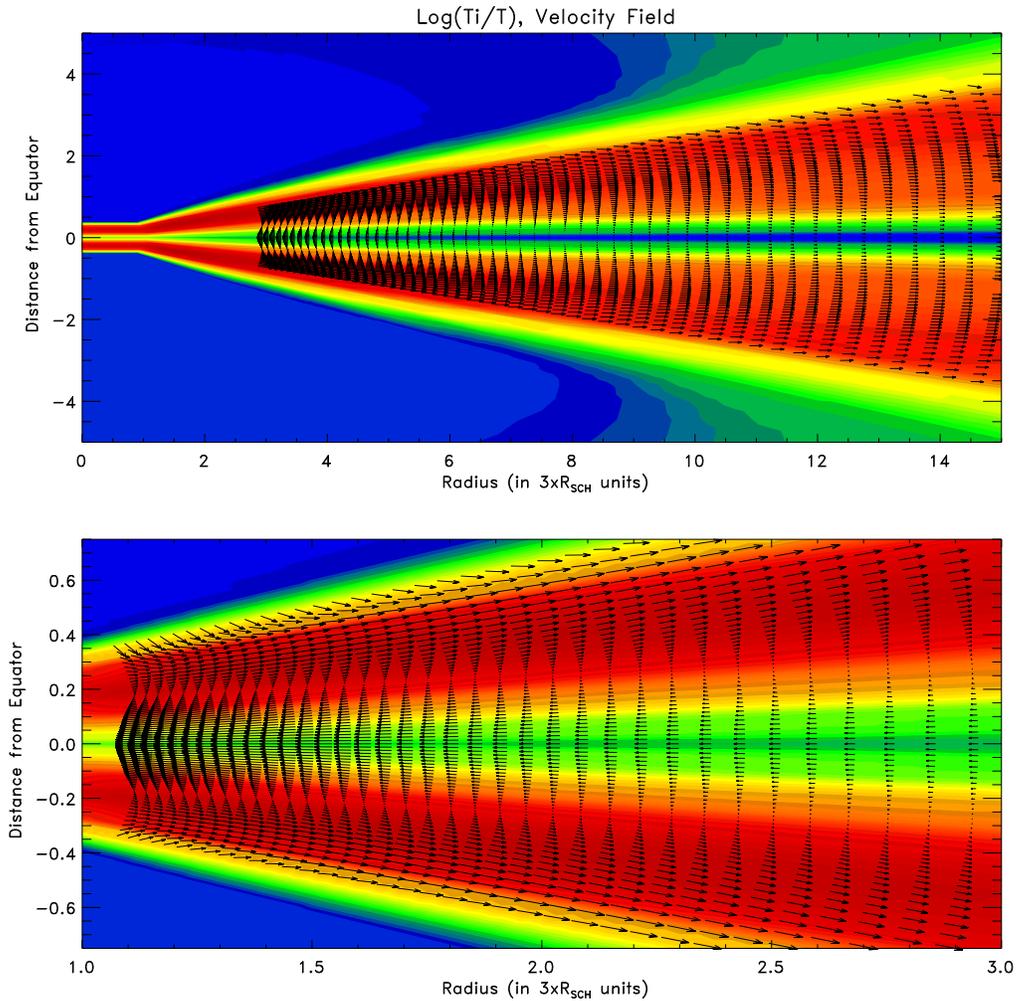}
}
\end{center}
{\vspace*{-0.4cm}}
\caption [ ] {The distribution of the velocity field superposed on the logarithmic-scaled 
        ratio of the ion- to electron-temperatures (red color corresponds to high ratios
           and blue to low-ratios).  The lower figure is a zoom-in of the flow
           configuration in the innermost part of the disk. 
  } 
\end{figure*}
\begin{figure}[htb]
\begin{center}
{\hspace*{-0.5cm}
\includegraphics*[width=9.5cm, bb=58 508 444 725,clip]
{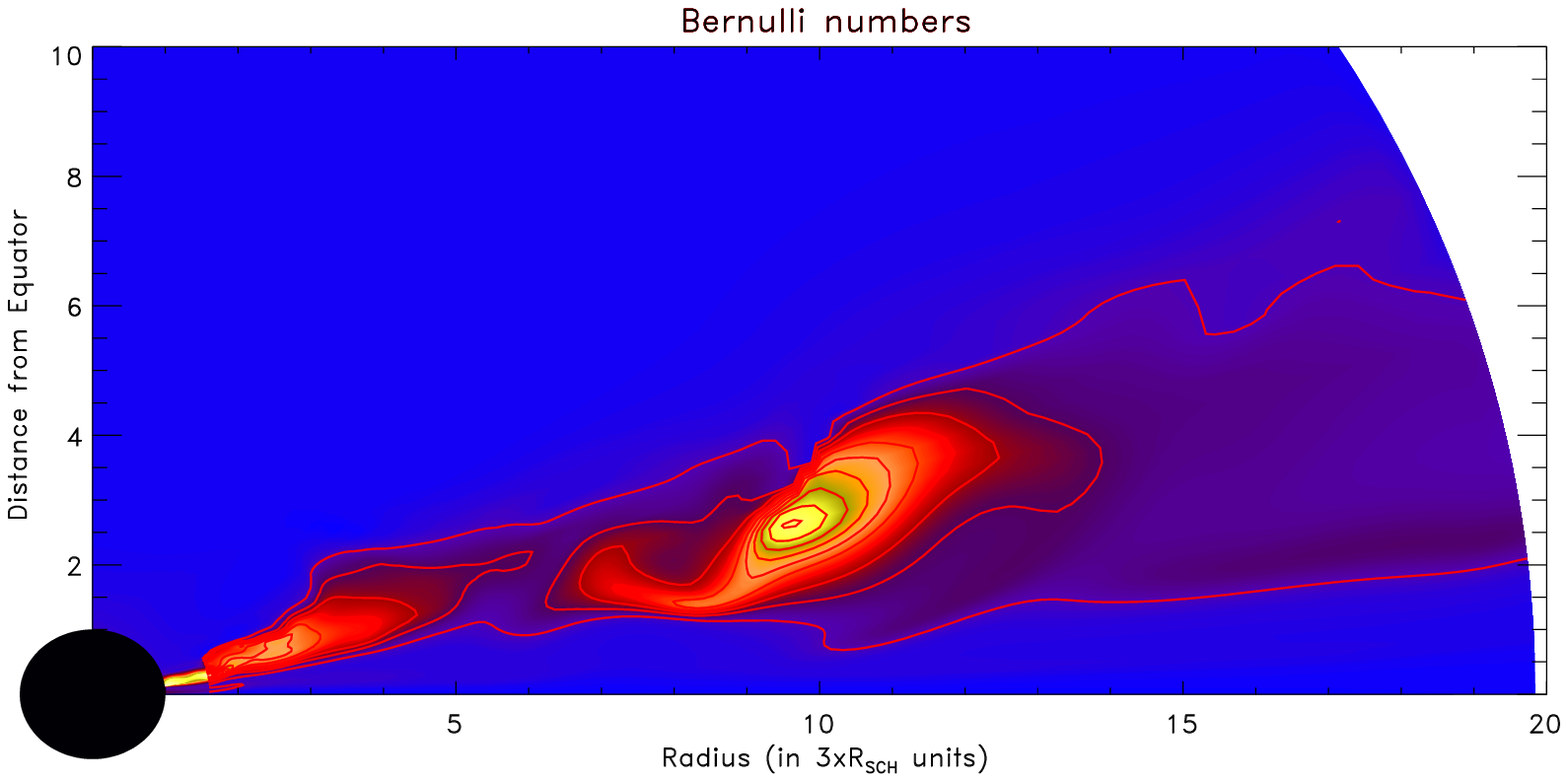}
}
\end{center}
{\vspace*{-0.4cm}}
\caption [ ] { A snap-shot of the distribution of the Bernoulli number in 
               two-dimensional time-dependent calculations. 
              The decrease from large to low positive values is represented
              via yellow, green and red colors.  The blue color corresponds
              to negative values.  The figure shows two ejected blobs of
              large positive energies in the TL. Ejection starts from the innermost
              region of the TL and evolves non-linearly.
  } 
\end{figure}
 
\section{Summary}

We have presented a multi-layer model for initiating ion-jets in AGNs.
The model agrees with a previous numerical study which confirmed 
the formation of thermally-induced outflows in the transition layer
(Hujeirat \& Camenzind 2000). Here we have shown that incorporating the effects of MFs
manifests their formation and dramatically strengthen their dynamics.
  
Three ingredients for initiating winds have been detected:
1) a highly diffusive plasma dominated by virial-hot ions, 
2) large scale magnetic fields that efficiently transport angular momentum from the disk
into the TL, where the plasma rotates super-Keplerian, and 
3) an underlying advection-dominated accretion disk. 

Taking into account that the corona is dynamically unstable, 
adopting a large scale magnetic topology, and allowing ion-electron
thermal decoupling appear to force accretion flows to undergo
a global energy re-distribution: confined inflows (negative
Bernoulli number) in the equatorial region and in the corona, and 
thermally and magneto-centrifugally-driven outflows in the TL 
characterized through a positive Bernoulli number. This feature may
survive under different conditions: strong MFs suppress turbulence, 
weakening thereby the effect of the turbulent-viscosity and dominate the 
transport of angular momentum. On the other hand weak MFs in rotating
stratified flows would be amplified via dynamo-action and reach equipartition,
beyond which turbulence is again suppressed. This interplay between MFs and 
turbulent-viscosity, Balbus-Hawley and Parker instabilities  may
settle into an equilibrium state, in which inflows are simultaneously
associated with low-cooling out-flowing ion-plasma.

We note that in the absence of thermal conduction and adopting 
the one-temperature description,  low-viscosity radiatively inefficient HD  
and MHD accretion flows become inevitably convection-dominated. Therefore, 
in the early phases of jet-initiation,
CDAFs may play an important role in powering the jets in AGNs and microquasars 
(Abramowicz et al. 2002). 
 
The multi-layer model presented here accommodates some elements of BP82.
In particular, we agree with BP82 about the necessity
for a super-Keplerian rotation of the plasma overlying the accretion
disk. However, the plasma here is dominated by highly-diffusive and virial-hot ions;
it does not require a special $\rm{B_\mathrm{P}}-$alignment with respect to the disk-normal 
to enable jet-launching, as ideal-MHD treatment requires. 
While our results agrees with the ion-torus model
with respect to the necessity of 2T-plasma to  maintain the ions hot for a
significant time of their propagation-life in the ISM, no signatures for the
formation of ion-supported tori have been detected (Rees et al. 1982).\\

Our results differ from ADAF and ADIOs in several issues, and in particular  with respect to
 1) the existence of a layer adjusting to the disk,
          where the plasma is found to rotate super-Keplerian, 
 2) the configurations of the in- and the out-flows,
 3) stability of the corona in the vicinity of the BH, 
 4) the transition from SS-disk to advection-dominated disks and 
 5) with respect to the essence of Bernoulli number in dissipative flows,
         i.e., a positive Bernoulli number is necessary but not sufficient
         for outflows (Abramowicz et al. 2000).\\

Finally, we note that since the flow in the TL is highly dissipative 
(strengthen thermal and rotational coupling with the central nucleus), 
the innermost region of the disk rotates synchronously with Kerr black holes
(due to the frame dragging effect), and since $\tau_\mathrm{rem}$ decreases 
with radius and depends inversely on $\rm{B_\mathrm{T}}$ and $\rm{B_\mathrm{P}}$, we think that
the plasma attached to the poloidal magnetic field
would be forced to deposit its angular momentum to the plasma in the TL, 
thereby considerably enhancing the centrifugal power and ejecting the ion-plasma
into space with relativistic speeds.

\end{document}